\documentclass{seg}
\usepackage{multirow}
\usepackage{enumitem}
\usepackage{pifont}
\usepackage{booktabs}

\title{ A Frequency-Aware Dynamic Knowledge Distillation Framework: An Effective Tool for Bridging Low- and High-Frequency Seismic Information}

\author{%
  Jun Ma$^{1}$\thanks{E-mail: majunacc@163.com} \and
  Xinyang Wang$^{2}$\thanks{E-mail: xywang18@126.com} \and
  Xintong Dong$^{1,*,}$\thanks{Corresponding author: 18186829038@163.com}
}

\date{1 September 2026}

\begin{document}

\maketitle

\begin{center}
    \footnotesize
    $^{1}$State Key Laboratory of Deep Earth Exploration and Imaging, College of Instrumentation and Electrical Engineering, Jilin University, Changchun, 130026, China.\\
    $^{2}$Key Laboratory of Modern Power System Simulation and Control \& Renewable Energy Technology, Ministry of Education (Northeast Electric Power University), Jilin 132012, China.\\
\end{center}

% --- Abstract ---
\begin{abstract}
Seismic data contain rich information across different frequency bands, with low-frequency components primarily characterizing large-scale geological structures and high-frequency components preserving fine-scale seismic details. Effectively integrating these frequency-dependent components is essential for seismic feature learning, so as to better preserve structural continuity and fine-scale details. Knowledge distillation provides an effective means for transferring informative representations from high-quality data. However, existing distillation-based frameworks usually treat seismic features in a full-band manner, ignoring relationships across frequency bands and thereby limiting the coordinated transfer of low- and high-frequency knowledge. To bridge low- and high-frequency seismic features through knowledge distillation, we propose a frequency-aware dynamic knowledge distillation framework (FADKD-Net), which establishes a teacher–student learning framework and performs frequency-aware knowledge transfer between low- and high- frequency bands. Specifically, FADKD-Net decomposes seismic features into low- and high-frequency components and performs targeted distillation to exploit their complementary information. Low-frequency distillation guides the student model to learn stable structural priors, thereby improving the overall continuity of seismic events. Meanwhile, high-frequency distillation enhances the detailed feature modeling and improves the representational capability for complex and small-scale structures. Furthermore, a cross-domain feature alignment strategy is proposed to reduce distributional discrepancies across different surveys and enhance the transferability of the seismic representations learned by FADKD-Net. We apply the proposed FADKD-Net to three-dimensional (3D) seismic data interpolation, and results on several field datasets demonstrate its ability to preserve the overall continuity of seismic events while accurately recovering fine-scale details. Furthermore, FADKD-Net exhibits robust generalization across different seismic surveys. \\
\textbf{Keywords:} Deep Learning, Seismics, Knowledge Distillation, Multi-Bands
\end{abstract}

% --- Introduction ---
\section{Introduction}
In recent years, deep learning (DL) methods have become increasingly important in seismic feature learning owing to its ability to learn complex nonlinear relationships directly from data \cite{ref16,ref19,ref17,ref18,ref45,ref20}, with successful applied to various seismic tasks, including interpolation \cite{ref21,ref29,ref22}, denoising \cite{ref46,ref47}, and super-resolution \cite{ref49,ref50}. However, most existing methods ignore the frequency characteristics of seismic data, without fully exploiting the rich information across different frequency bands. In seismic data, low-frequency components mainly characterize large-scale geological structures and provide stable structural priors, whereas high-frequency components preserve fine-scale seismic details and local geological details. Effectively exploiting information across different frequency bands is essential for learning comprehensive seismic features. Motivated by this property, several studies have begun to use the multi-band information in seismic applications. For example, \cite{ref52} propose a pattern-based method for adaptive multiple subtraction that learns the characteristic patterns of multiples across different frequency bands, thereby improving multiple-suppression accuracy. \cite{ref53} introduce a generative dual-band learning strategy for low-frequency extrapolation, which facilitates knowledge transfer between synthetic and field seismic data and alleviates cycle-skipping problems in full-waveform inversion. \cite{ref54} further develop a multi-band learning strategy for seismic interpolation, in which different frequency bands are processed independently to mitigate inter-band interference and improve interpolation accuracy. These studies indicate that different frequency components play different but complementary roles in seismic information. However, effectively transferring and jointly exploiting these complementary frequency information remains challenging, particularly when the available observations are degraded or acquired from different seismic surveys.

Knowledge distillation is an effective knowledge transfer strategy that plays an important role in seismic data processing. The core concept is to employ a teacher model with larger capacity and stronger representation ability to guide a compact, computationally efficient student model \cite{ref35}. In addition to learning from ground-truth data, the student can exploit the teacher's output and intermediate features, thereby acquiring informative prior knowledge learned from high-quality data. By introducing the teacher model, the distillation framework transfers implicit prior knowledge to the student model, enhancing the student model’s performance and generalization ability across different data distributions. Owing to these advantages, knowledge distillation has been increasingly applied to various seismic data processing tasks, including denoising \cite{ref41, ref56}, fault detection \cite{ref36, ref39, ref58}, image super-resolution \cite{ref37, ref40, ref42}, full waveform inversion \cite{ref38}, subsurface imaging \cite{ref55}, and so on\cite{ref59, ref60, ref61}. These studies demonstrate the potential of knowledge distillation for transferring seismic knowledge and improving model performance under complex geological conditions. However, existing seismic knowledge distillation frameworks usually focus on transferring seismic knowledge in a full-band manner, without explicitly considering the frequency-dependent characteristics of seismic features. Treating them as undifferentiated full-band features may overlook the relationships across frequency bands, thereby limiting the coordinated transfer of low- and high-frequency knowledge. In particular, indiscriminate knowledge transfer may fail to provide the student model with targeted structural guidance from low-frequency information while simultaneously exploiting fine-scale information from high-frequency components.

To address these challenges, we propose a frequency-aware dynamic knowledge distillation framework (FADKD-Net) for seismic feature learning. FADKD-Net establishes a teacher–student learning framework in which the teacher model extracts high-quality prior knowledge from high-quality data and guides the student model to learn from degraded observations. To enable frequency-specific knowledge transfer, a frequency-aware distillation strategy is introduced by decomposing seismic features into low- and high-frequency components. Specifically, low-frequency distillation guides the student model to learn stable structural priors, improving the overall continuity of seismic events. High-frequency distillation enhances fine-scale feature modeling and improves the representation of complex and small-scale structures. Furthermore, a cross-domain feature alignment strategy is incorporated to reduce distributional discrepancies across different seismic surveys. To evaluate the effectiveness of the proposed framework, 3D seismic interpolation is selected as a downstream task. Experiments on several field seismic datasets demonstrate that FADKD-Net achieves promising performance and robust generalization across different surveys. The main contributions of this work are summarized as follows:

(1)	We propose a frequency-aware dynamic knowledge distillation framework for seismic feature learning. By establishing a teacher–student architecture, the framework transfers prior knowledge from high-quality data to guide feature learning from degraded observations.

(2)	We develop a frequency-aware distillation strategy that decomposes features into low- and high-frequency components. Low-frequency distillation provides stable structural priors to preserve seismic event continuity, whereas high-frequency distillation facilitates fine-scale feature modeling and the recovery of detailed geological information.

(3)	We introduce a cross-domain feature alignment strategy to alleviate distribution discrepancies across different seismic surveys, thereby improving the transferability and generalization of the learned seismic features.

(4)	We validate the effectiveness of the proposed framework on multiple field datasets using 3D interpolation as a downstream task. Experimental results demonstrate that FADKD-Net achieves promising performance and strong generalization ability across different surveys.

\begin{figure}[htpb]
	\centering
	{\includegraphics[width=0.99\linewidth]{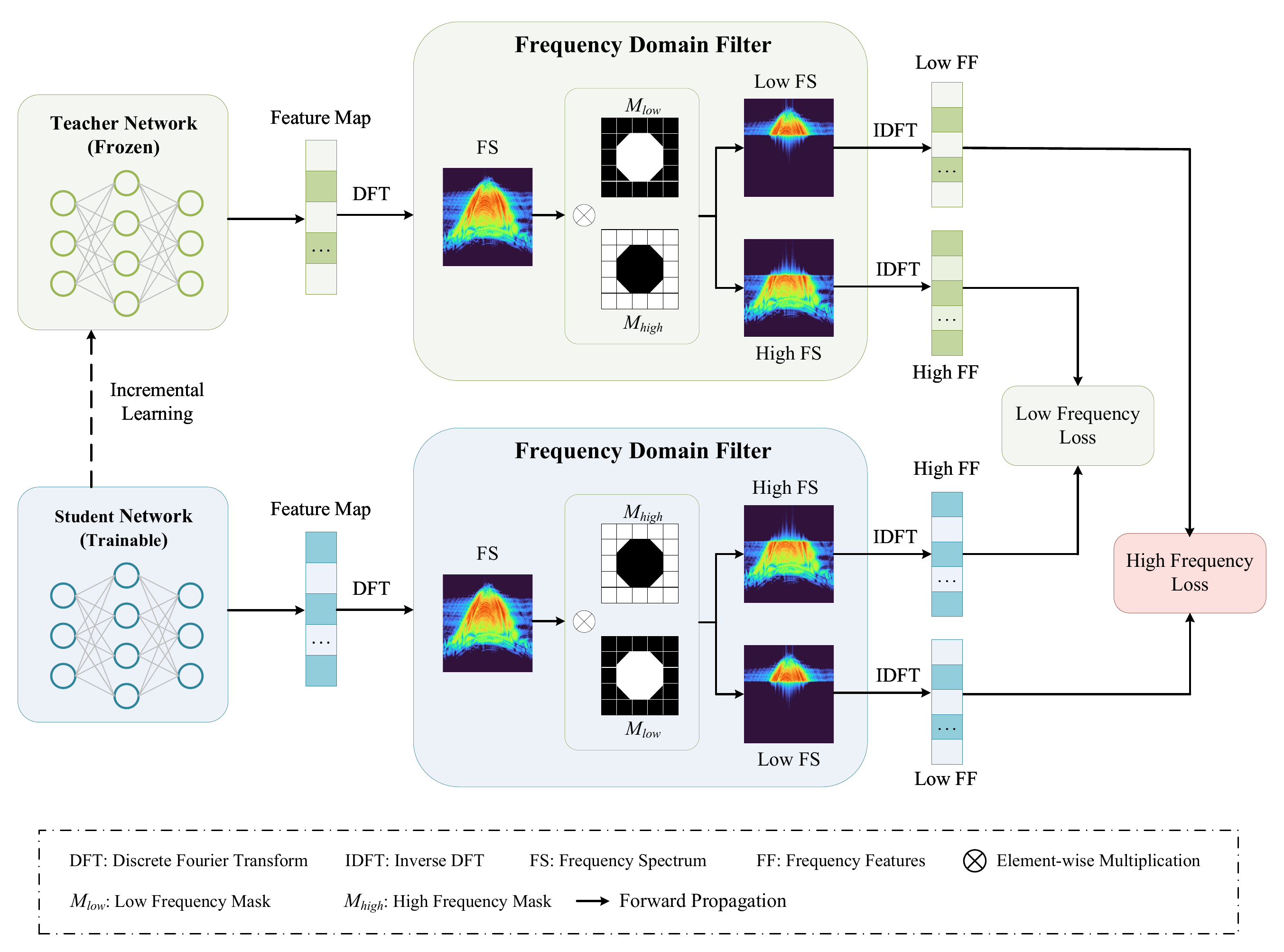}}
	\caption{The frequency-aware dynamic knowledge distillation framework.}
	\label{fig_framework}
\end{figure}

\begin{figure}[htpb]
	\centering
	{\includegraphics[width=0.99\linewidth]{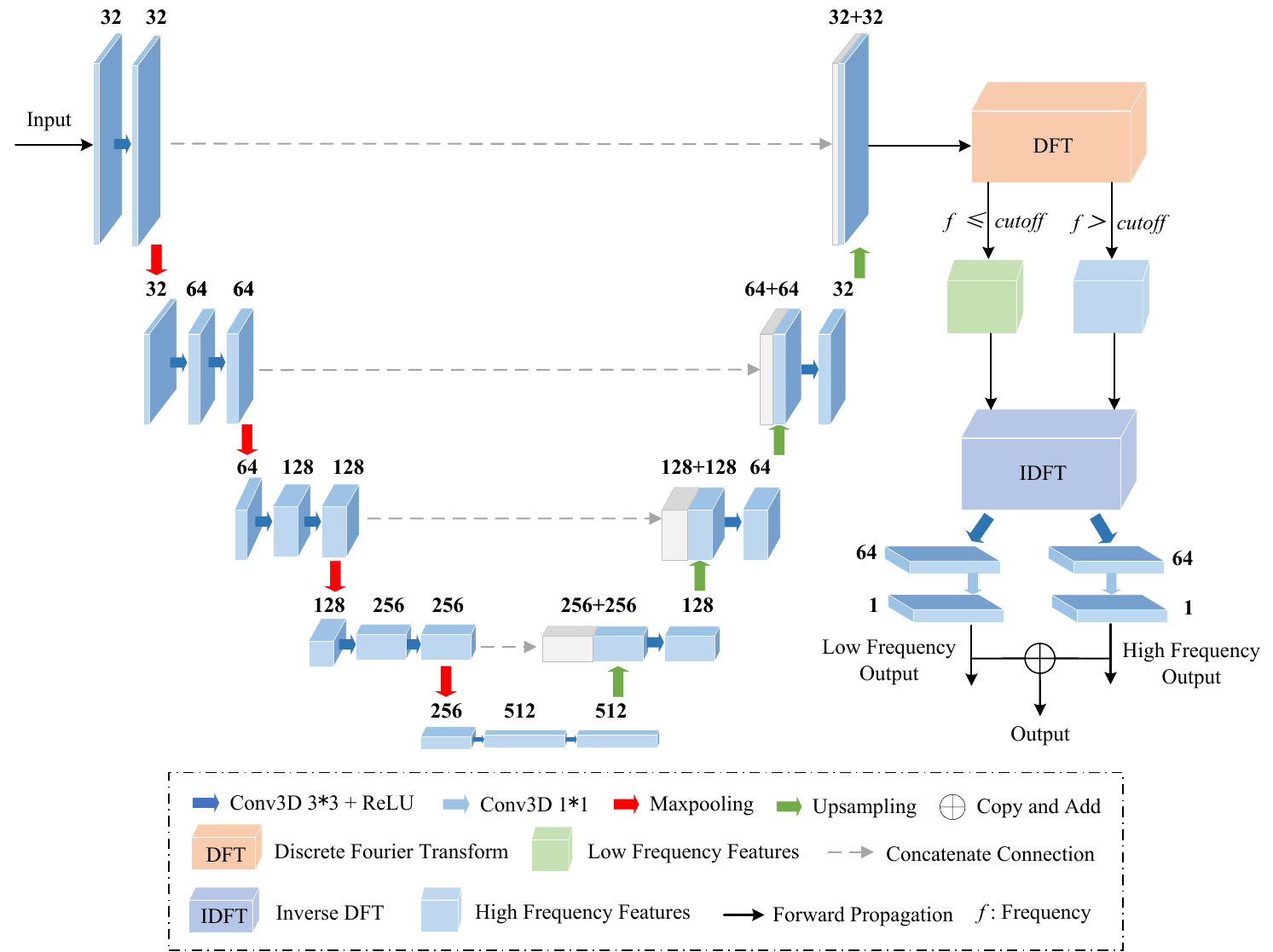}}
	\caption{The multi-branch U-Net model.}
	\label{fig_unet}
\end{figure}

% --- Methods ---
\section{Methodology}

In this section, we propose a frequency-aware dynamic knowledge distillation framework, as illustrated in Figure~\ref{fig_framework}. Specifically, full-band features are decomposed into low-frequency structural components and high-frequency detailed components. Based on this frequency decomposition, a teacher–student framework is established to achieve frequency-specific knowledge distillation, enabling differentiated transfer of structural and detail information. Furthermore, cross-domain feature alignment is performed in the frequency domain to alleviate distribution discrepancies across different datasets. The detailed implementations of each component are described in the following section

\subsection{Network Architecture}
In this study, we employ a 3D U-Net as the backbone network for seismic feature learning \cite{ref43}. The U-Net architecture exhibits strong capability in multi-scale feature extraction and adopts a symmetric encoder–decoder structure with skip connections. The encoder progressively downsamples the input data to extract deep multi-scale features, while the decoder gradually restores spatial resolution via upsampling to reconstruct the output. Skip connections are introduced to integrate shallow encoder features with corresponding decoder features at the same resolution level. This design compensates for information loss caused by downsampling and effectively preserves structural details in 3D seismic volumes.

The overall network architecture is illustrated in Figure~\ref{fig_unet}. Each encoder block consists of two $3 \times 3 \times 3$ convolution layers followed by a ReLU activation, and a max-pooling layer is subsequently applied to reduce feature map resolution while doubling the number of channels from 32 to 512. This hierarchical feature extraction mechanism enables deeper layers to capture global structural patterns of large-scale geological features, while shallow layers retain rich local details. In the decoder, trilinear interpolation is employed to recover the spatial resolution of feature maps, and skip connections are used to integrate encoder and decoder features. Unlike conventional single-output architectures, we introduce a multi-branch prediction head at the network output to facilitate frequency-aware feature modeling. Specifically, the extracted features are transformed into the frequency domain using the Fourier transform and decomposed into low- and high-frequency bands for independent modeling. This design establishes the basis for subsequent frequency-decomposed knowledge distillation.

\subsection{Frequency-decoupled Learning}
Seismic signals exhibit distinct physical characteristics across different frequency bands. The low-frequency components mainly characterize large-scale subsurface structures and remain relatively stable across different surveys and acquisition scenarios. High-frequency components contain detailed geological information, such as faults and stratigraphic variations, but are more sensitive to noise and local geological conditions, resulting in higher uncertainty and stronger data dependence. Existing DL-based methods usually adopt full-band unified modeling strategies, ignoring the non-uniform distribution of frequency-domain features. Such indiscriminate learning makes it difficult to simultaneously preserve structural continuity and recover high-frequency details, often leading to over-smoothed structures or missing fine-scale information. Therefore, we introduce a frequency-decoupled representation strategy to enable adaptive and differentiated modeling of different frequency components.

To achieve adaptive frequency decomposition, we perform trace-wise Fourier transforms (FT) on the seismic data and accumulate the corresponding energy spectra to construct the energy–frequency curve. The cutoff frequency ${f_c}$ is determined according to cumulative spectral energy rather than predefined frequency values, allowing adaptive decomposition across different seismic surveys. Specifically, ${f_c}$ is selected as the frequency point where the cumulative energy reaches one-third of the total energy, while the remaining two-thirds of the spectral energy are assigned to the high-frequency component. Given an input seismic volume $X \in {\mathbb{R}^{T \times X \times Y}}$, where T denotes the time dimension and $X$, $Y$ represent the two spatial dimensions, the Discrete Fourier transform (DFT) is applied along the temporal axis to transform the signal into the frequency domain, producing its spectral representation:
\begin{equation}
	\label{eq1}
	{X_f} = DFT(X).
\end{equation}
Then, we define a low-pass filter ${M_{low}}(f)$ and a high-pass filter ${M_{high}}(f)$ to decompose ${X_f}$ into low-frequency $X_{low}^f$ and high-frequency bands $X_{high}^f$, where:
\begin{equation}
	\label{eq2}
	{M_{low}}(f) = \left\{ {\begin{array}{*{20}{l}}
			{1,}&{|f| \le {f_c}}\\
			{0,}&{|f| > {f_c}}
	\end{array}}, \right.{M_{high}}(f) = 1 - {M_{low}}(f),
\end{equation}
\begin{equation}
	\label{eq3}
	X_{low}^f = {X^f} \times {M_{low}}(f),X_{high}^f = {X^f} \times {M_{high}}(f).
\end{equation}
After frequency domain filtering, we apply the Inverse Discrete Fourier transform (IDFT) to map $X_{low}^f$ and $X_{high}^f$ back into the spatial domain, yielding the decoupled low-frequency feature ${X_{low}}$ and high-frequency feature ${X_{high}}$, respectively:
\begin{equation}
	\label{eq4}
	{X_{low}} = IDFT(X_{low}^f),{X_{high}} = IDFT(X_{high}^f).
\end{equation}
Through the above process, the full-band processing task is decomposed into frequency-specific sub-tasks: low-frequency structural recovery and high-frequency detail reconstruction. This frequency-decoupled representation not only enhances the model’s feature representation capability but also establishes a foundation for subsequent frequency-decomposed knowledge distillation.

\subsection{Dynamic Knowledge Distillation }
In practical seismic processing scenarios, data distributions often exhibit significant variations across different surveys, which may cause noticeable performance degradation when models trained in the source domain are directly applied to the target domain. Moreover, target-domain seismic data are often limited and degraded, making it difficult for models to learn reliable seismic features solely from the available observations.

The teacher model is trained on high-quality source domain data to acquire seismic knowledge, which serves as prior guidance for the student model. The student model is subsequently trained on degraded target domain data under the supervision of the teacher, enabling the transfer of learned knowledge across domains. During optimization, the parameters of the student model are iteratively updated via backpropagation and gradient descent:
\begin{equation}
	\label{eq5}
	\theta _S^{(t + 1)} = \theta _S^{(t)} - \eta  \cdot {\nabla _{{\theta _S}}}{L_{total}},
\end{equation}
where $\eta $ denotes the learning rate, and ${\nabla _{{\theta _{\cal S}}}}{L_{total}}$ represents the gradient of the total loss with respect to ${\theta _S}$. It is worth noting that the teacher model does not participate in backpropagation. Instead, its parameters ${\theta _T}$ are updated only through exponential moving average (EMA):
\begin{equation}
	\label{eq6}
	\theta _T^{(t)} = \alpha \theta _T^{(t - 1)} + (1 - \alpha )\theta _S^{(t)},
\end{equation}
where ${\alpha _t} = \min (1 - 1/(t + 1),{\alpha _{max}})$ is the smoothing coefficient, which is dynamically adjusted according to the global training step $t$. Unlike conventional knowledge distillation strategies that employ a fixed teacher model, the proposed dynamic distillation mechanism enables continuous adaptation of the teacher model throughout training. Specifically, the teacher model is updated through an EMA of the student parameters, allowing it to maintain stable high-fidelity structural priors while gradually incorporating target domain characteristics captured by the student model. This dynamic update establishes a bidirectional knowledge evolution process between the teacher and student models, where the teacher provides reliable prior guidance, and the student feedback further improves domain adaptability. Since the teacher optimization is independent of gradient backpropagation, the proposed strategy avoids unstable parameter oscillations caused by direct gradient-based teacher updating and enables more robust knowledge transfer across different surveys.

\subsection{Frequency-Aware Cross-Domain Feature Alignment}
Although the distillation mechanism improves model performance, considerable domain shift still exists across different surveys, particularly in complex geological environments and high-frequency components. To further enhance cross-domain generalization, we introduce a frequency-aware feature alignment strategy that performs distribution alignment separately on low- and high-frequency features in the frequency domain. Specifically, to reduce amplitude scale discrepancies across different surveys, the decoupled frequency components are normalized before calculating the alignment loss:
\begin{equation}
	\label{eq7}
	\widehat X = \frac{{X - \mu }}{\sigma },
\end{equation}
where $\mu $ and $\sigma $ denote the mean and standard deviation of the corresponding feature space, respectively. This normalization operation ensures that knowledge transfer between the teacher and student models is performed under a consistent numerical distribution.

Low-frequency features $X_{low}$ mainly characterize the continuity of geological structures. Therefore, a cosine similarity loss is employed to constrain the low-frequency feature representations. By measuring the angular discrepancy between flattened feature vectors, the proposed strategy preserves directional consistency and lateral continuity of geological structures:
\begin{equation}
	\label{eq8}
	{L_{low}} = 1 - \frac{1}{N}\sum\limits_{i = 1}^N {\frac{{{\rm{vec}}{{(X_{low,i}^{stu})}^T}{\rm{vec}}(X_{low,i}^{tea})}}{{\left \|{\rm{vec}}(X_{low,i}^{stu}) \right \|_2\left \|{\rm{vec}}(X_{low,i}^{tea}) \right \|_2}}} ,
\end{equation}
where ${\rm{vec}}( \cdot )$ denotes the vectorization operator, $N$ represents the batch size, $\left \| \cdot \right \|{_2}$ denotes the ${L_2}$ norm, and $X_{low,i}^{stu}$ and $X_{low,i}^{tea}$ are the low-frequency feature maps of the $i-th$ sample from the student and teacher networks, respectively.

High-frequency features ${X_{high}}$ represent rapidly varying components containing rich local details. To improve the robustness of the model to these features, we introduce a logarithmic mean squared error (Log-MSE) loss. This formulation compresses the dynamic range of feature errors, encouraging the model to focus on relative discrepancies rather than absolute magnitudes. Consequently, it preserves accurate reconstruction of strong signals while enhancing the recovery of fine-scale details. Its mathematical formulation is given as follows:
\begin{equation}
	\label{eq9}
	{L_{high}} = \frac{1}{N}\sum\limits_{i = 1}^N {\log } \left( {1 + {{\left\| {X_{high,i}^{stu} - X_{high,i}^{tea}} \right\|}_2}} \right).
\end{equation}
To maintain spatial-domain consistency and data fidelity of the reconstructed data, a global reconstruction loss is further introduced:
\begin{equation}
	\label{eq10}
	{L_{rec}} = \left \|X_{out}^{stu} - {X_{gt}} \right \|_2,
\end{equation}
where $X_{out}^{stu} = X_{low}^{stu} + X_{high}^{stu}$ denotes the reconstruction output of the student network, ${X_{gt}}$ represents the ground truth. The overall optimization objective is formulated as:
\begin{equation}
	\label{eq11}
	{L_{total}} = {\lambda _1}{L_{rec}} + {\lambda _{low}}{L_{low}} + {\lambda _{high}}{L_{high}},
\end{equation}
where ${\lambda _1} = 1,{\lambda _{low}} = 0.1,{\lambda _{high}} = 0.2$ is the weighting coefficient determined via grid search. By incorporating frequency-domain alignment, the proposed framework effectively alleviates discrepancies across different surveys and improves cross-domain generalization capability.

\section{Results}
In this section, we present the experimental settings, including implementation details, training configurations, and evaluation metrics. Seismic interpolation is adopted as the validation task. Multiple field seismic datasets are employed to further assess its reconstruction performance and generalization capability.

\subsection{Implementation Details}
The proposed method is implemented using the PyTorch framework and trained on an Nvidia GeForce RTX 4090D GPU with 24 GB of memory. During training, the Adam optimizer is adopted with an initial learning rate of 10-4. The batch size is set to 16, and the model is trained for 40-100 epochs. The proposed method is trained in a supervised manner, where the complete seismic volumes are used as labels. Training samples are extracted with a size $64 \times 64 \times 64$, and the corresponding inputs are generated by randomly masking 30\%–70\% of the traces from the complete data. The model achieving the highest Peak Signal-to-Noise Ratio (PSNR) on the validation set is selected for testing.

To quantitatively evaluate interpolation performance, two widely used metrics, PSNR and Structural Similarity Index Measure (SSIM), are adopted. The PSNR is defined as:
\begin{equation}
	\label{eq12}
	PSNR = 10 \cdot {\log _{10}}(\frac{{Max{{({y_{gt}})}^2}}}{{MSE({y_{gt}},{y_{pred}})}}),
\end{equation}
where ${y_{gt}}$ and ${y_{pred}}$ denote the ground truth and reconstructed values, respectively, $Max( \cdot )$ denotes the maximum possible pixel value of the data, and $MSE( \cdot )$ is the Mean Squared Error. A higher PSNR value indicates better reconstruction quality. The SSIM is formulated as:
\begin{equation}
	\label{eq13}
	SSIM = \frac{{(2{\mu _{{y_{gt}}}}{\mu _{{y_{pred}}}} + {c_1})(2{\sigma _{{y_{gt}}}}_{{y_{pred}}} + {c_2})}}{{(\mu _{{y_{gt}}}^2 + \mu _{{y_{pred}}}^2 + {c_1})(\sigma _{{y_{gt}}}^2 + \sigma _{{y_{pred}}}^2 + {c_2})}},
\end{equation}
where ${\mu _{{y_{gt}}}}$ and ${\mu _{{y_{pred}}}}$ denote the mean values,  $\sigma _{{y_{gt}}}^2$ and $\sigma _{{y_{pred}}}^2$ represent the variances, and ${\sigma _{{y_{gt}}{y_{pred}}}}$ is the covariance between ${y_{gt}}$ and ${y_{pred}}$, ${c_1}$ and ${c_2}$ are small constants to stabilize the division. The SSIM value ranges from 0 to 1, where values closer to 1 indicate stronger structural similarity between the reconstructed data and the ground truth.

\subsection{Test on Mobil AVO Viking Graben Line 12}
To evaluate the performance of FADKD-Net, we compare it with a traditional interpolation method, damped rank-reduction (DRR) \cite{ref44}, and a U-Net model trained only on target domain data, which is regarded as the baseline. The Mobil AVO Viking Graben Line 12 (MAVGL 12) dataset is a 2D prestack field dataset with a sampling interval of 4 ms. Following previous studies \cite{ref26, ref43}, the original data are reorganized into a 3D volume for interpolation experiments. From this volume, a sub-volume with dimensions of $480 \times 864 \times 96$ ($Time \times Crossline \times Inline$) is extracted for training. Using a sliding window of size $64 \times 64 \times 64$ with a stride of $32 \times 32 \times 32$, 728 patches are generated and divided into training and validation sets with a ratio of 9:1. In addition, an independent testing block with dimensions of $224 \times 128 \times 96$, which is excluded from the training region, is used for evaluation. 

\begin{figure}[htpb]
	\centering
	{\includegraphics[width=0.99\linewidth]{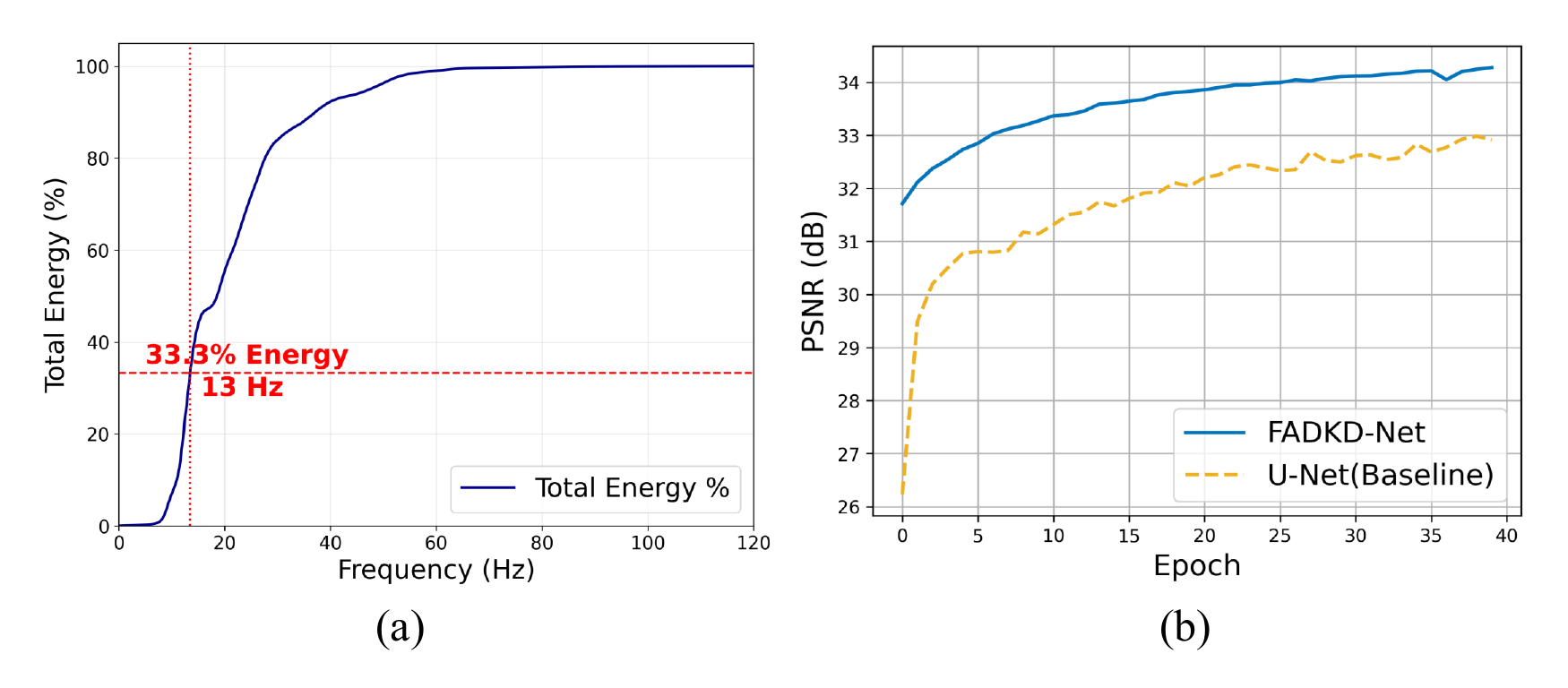}}
	\caption{Frequency-energy spectrum and PSNR evaluation on the MAVGL 12 dataset.  (a) The frequency-energy spectrum; (b) PSNR values on the validation sets.}
	\label{fig_viking_dataset}
\end{figure}

Due to the limited observations and complex geological structures of MAVGL 12, directly training an end-to-end model provides insufficient supervision for accurate reconstruction. To address this limitation, we adopt a teacher–student learning framework to incorporate prior knowledge into the reconstruction process. Specifically, the teacher model is trained using synthetic prestack seismic data from the Sandia/SEG Salt Model 45, with a cutoff frequency of 18 Hz for frequency decomposition. A seismic volume with dimensions of $416 \times 192 \times 808$ ($Time \times Crossline \times Inline$) is divided into $64 \times 64 \times 64$ patches using a sliding window with a stride of $32 \times 32 \times 32$, resulting in 1440 samples for the teacher model. Through this training process, the teacher model learns a reliable mapping between under-sampled data and fully sampled data. Subsequently, the student model is trained on the MAVGL 12 dataset under the guidance of the teacher model. Rather than relying solely on limited target domain supervision, the teacher–student framework enables the transfer of high-quality prior knowledge learned from synthetic data, thereby guiding the student to recover missing traces more effectively. 

\begin{figure}[htpb]
	\centering
	{\includegraphics[width=0.99\linewidth]{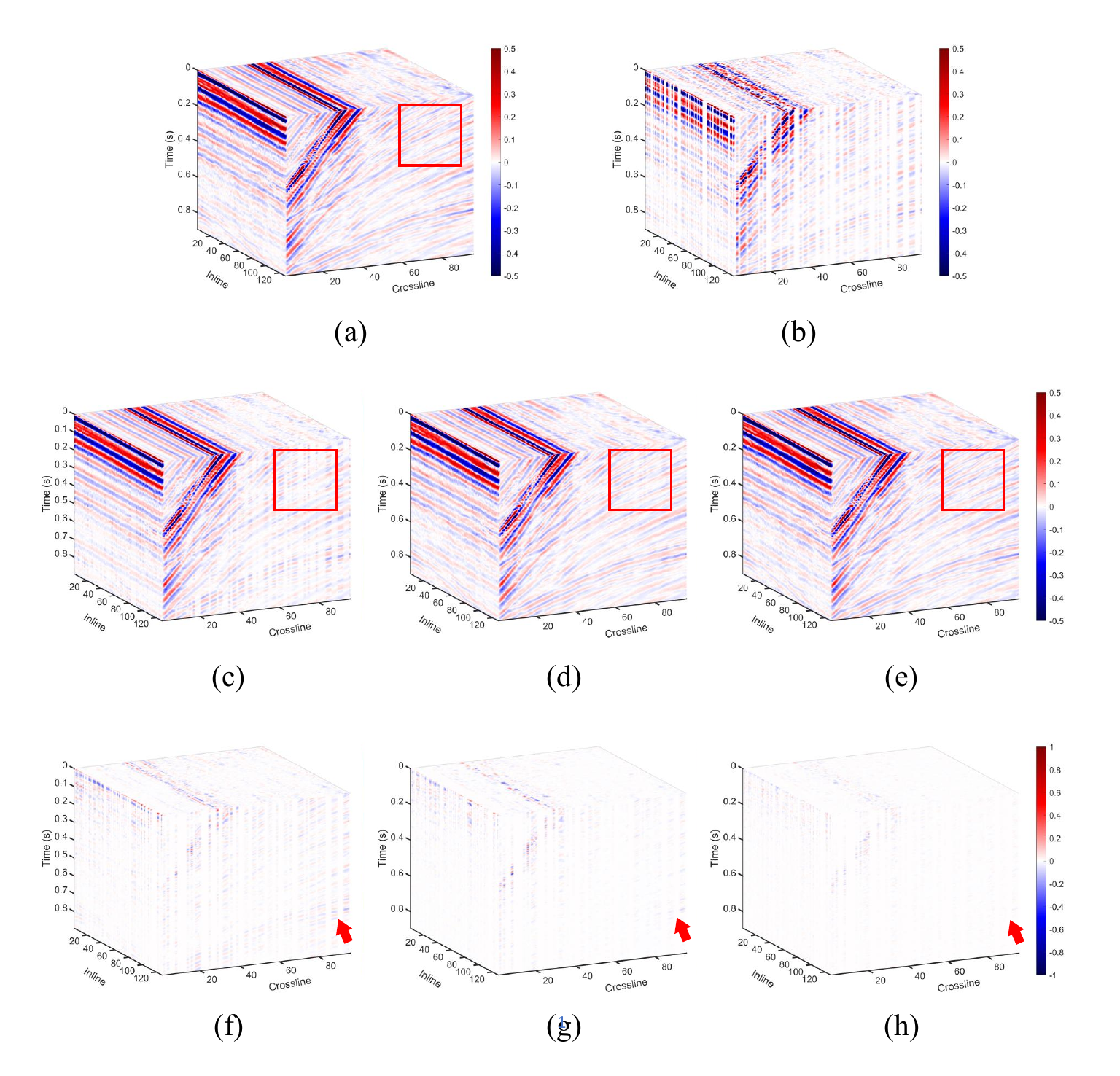}}
	\caption{Interpolation results on the MAVGL 12 dataset. (a) Ground truth. (b) Data with 50\% randomly sampled. (c)-(e) Interpolation results corresponding to the DRR, U-Net(Baseline), and FADKD-Net, respectively. (f)–(h) Residual between (a) and (c)-(e), respectively.}
	\label{fig_viking_results}
\end{figure}

Figure~\ref{fig_viking_dataset}(a) presents the frequency–energy spectrum of the MAVGL 12 dataset. Based on the energy distribution, the frequency point corresponding to one-third of the cumulative total energy (13 Hz) is selected as the cutoff frequency. Frequencies below this threshold are defined as low-frequency components, while the remaining frequencies are regarded as high-frequency components. Figure~\ref{fig_viking_dataset}(b) shows the PSNR curves on the validation set during training. It can be observed that FADKD-Net consistently achieves higher PSNR values throughout the training process, demonstrating its improved interpolation capability. Figure~\ref{fig_viking_results} shows the interpolation results of different methods. The reconstructed results obtained by DRR and U-Net(Baseline) appear visually similar to the ground truth; however, noticeable differences can still be observed in the residual map (see red arrows), indicating poor reconstruction quality (see the red box area). In contrast, with the incorporation of the frequency-aware dynamic knowledge distillation framework, FADKD-Net effectively exploits prior knowledge and achieves improved interpolation performance. The reconstructed seismic traces exhibit better continuity and structural consistency, while the residual maps show the lowest signal leakage across all compared methods. Quantitative metrics: DRR achieves a PSNR of 39.48 dB and an SSIM of 0.9529, while U-Net (Baseline) obtains a PSNR of 41.49 dB and an SSIM of 0.9810. In comparison, FADKD-Net achieves the best reconstruction performance, reaching a PSNR of 45.18 dB and an SSIM of 0.9896. These results demonstrate that the proposed framework provides more accurate seismic interpolation by effectively combining target-domain information with prior knowledge from the teacher model. 

\begin{figure}[htpb]
	\centering
	{\includegraphics[width=0.99\linewidth]{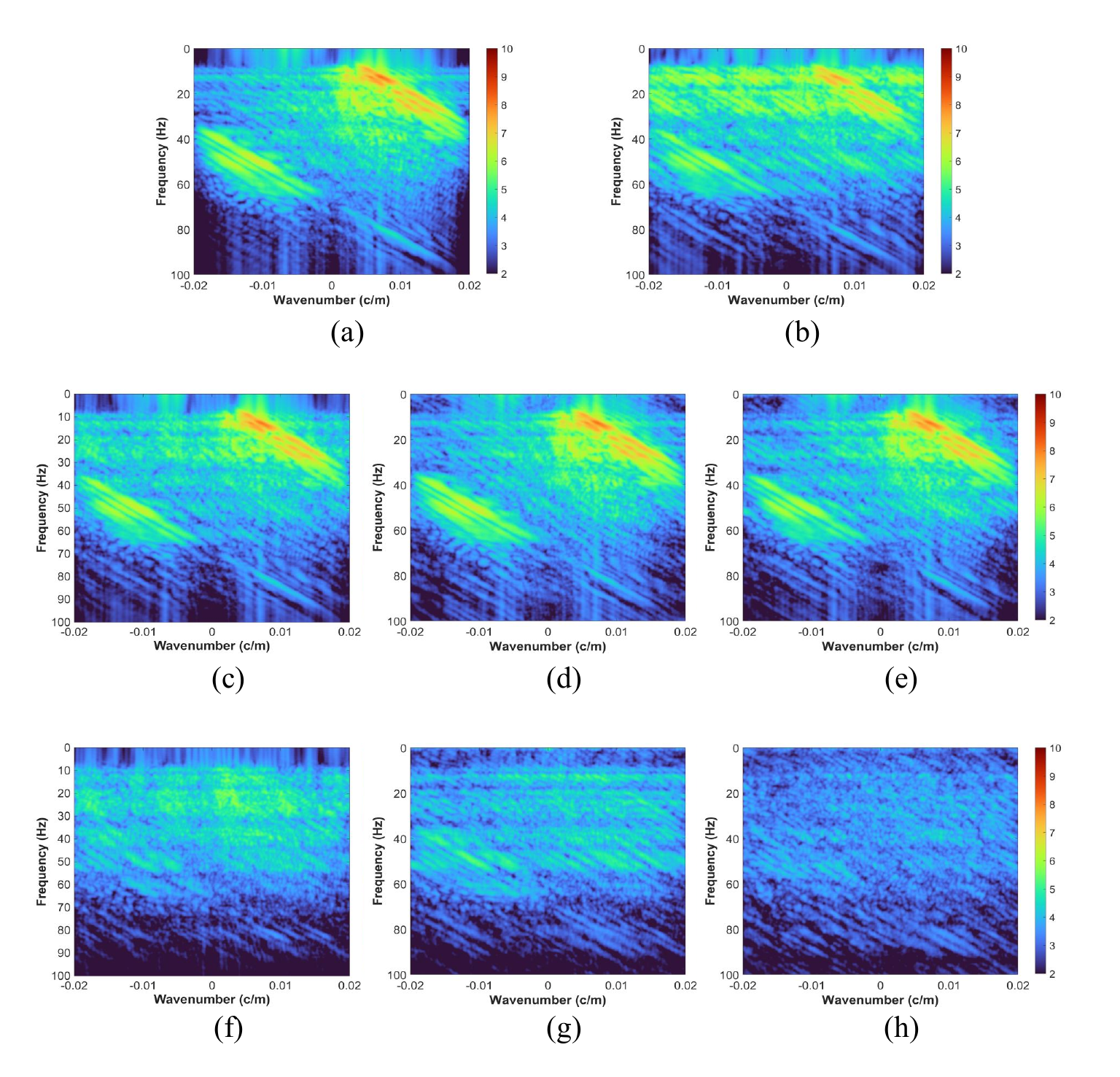}}
	\caption{F-K spectra comparisons of the MAVGL 12 dataset. (a)-(h) F-K spectra of Figure~\ref{fig_viking_results}(a)-(h).}
	\label{fig_viking_fk}
\end{figure}

\begin{figure}[htpb]
	\centering
	{\includegraphics[width=0.99\linewidth]{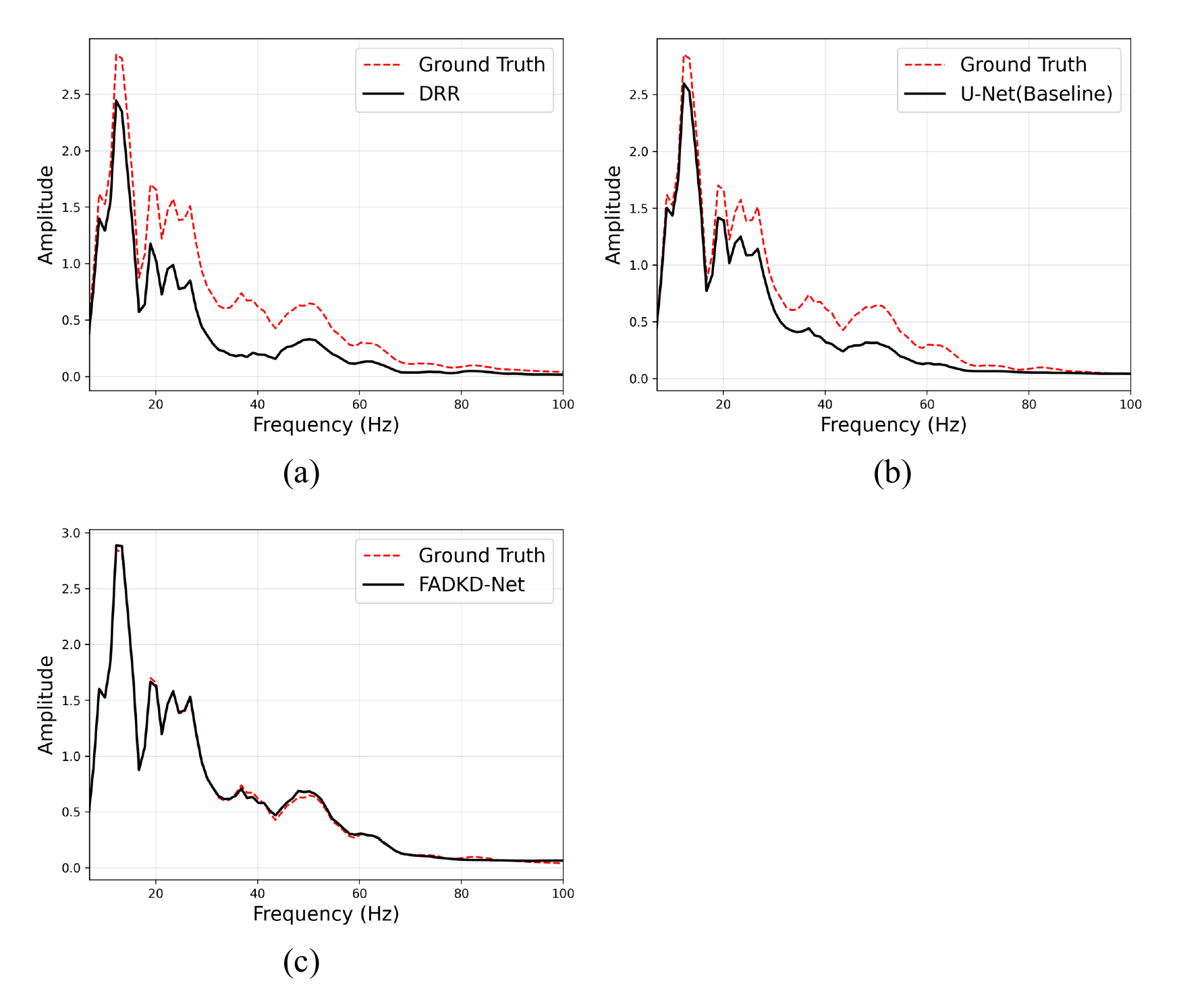}}
	\caption{Spectrum comparisons of the MAVGL 12 dataset. (a)-(c) Spectrum results corresponding to the DRR, U-Net(Baseline), and FADKD-Net, respectively.}
	\label{fig_viking_freq}
\end{figure}

For further frequency domain analysis, Figure~\ref{fig_viking_fk} presents the FK spectra corresponding to the results in Figure~\ref{fig_viking_results}. Compared with DRR and U-Net (Baseline), FADKD-Net exhibits reduced signal leakage and better preservation of frequency domain characteristics. Figure~\ref{fig_viking_freq} further provides a comparison of amplitude spectra. It can be observed that FADKD-Net achieves a closer match to the ground truth spectrum across the full frequency band range, indicating its ability to reconstruct both low-frequency structural information and high-frequency details more accurately. 

Overall, by introducing the teacher–student framework and frequency-aware distillation strategy, FADKD-Net achieves improved interpolation performance in both temporal and frequency domains. The teacher model provides reliable prior knowledge learned from synthetic data, while the student model adapts this knowledge to the limited MAVGL 12 dataset. This cross-domain knowledge transfer effectively reduces the discrepancy between synthetic and field seismic data, leading to enhanced reconstruction accuracy and improved generalization capability.

\subsection{Test on Kerry}
The Kerry dataset is a poststack field dataset acquired from New Zealand with a sampling interval of 4 ms. It contains complex subsurface structures and strong geological variability, providing a challenging test case for evaluating the generalization capability of interpolation methods. From a seismic volume with dimensions of $352 \times 192 \times 480$ ($Time \times Crossline \times Inline$), 1920 patches with a size $64 \times 64 \times 64$ are extracted using a sliding window with a stride of $32 \times 32 \times 32$. These samples are randomly divided into training and validation sets with a ratio of 9:1. In addition, an independent testing volume with dimensions of $224 \times 192 \times 64$, which is excluded from the training region, is used for testing. Similar to the MAVGL 12 experiment, a teacher–student learning framework is adopted to introduce prior knowledge into the reconstruction process. 

\begin{figure}[htpb]
	\centering
	{\includegraphics[width=0.99\linewidth]{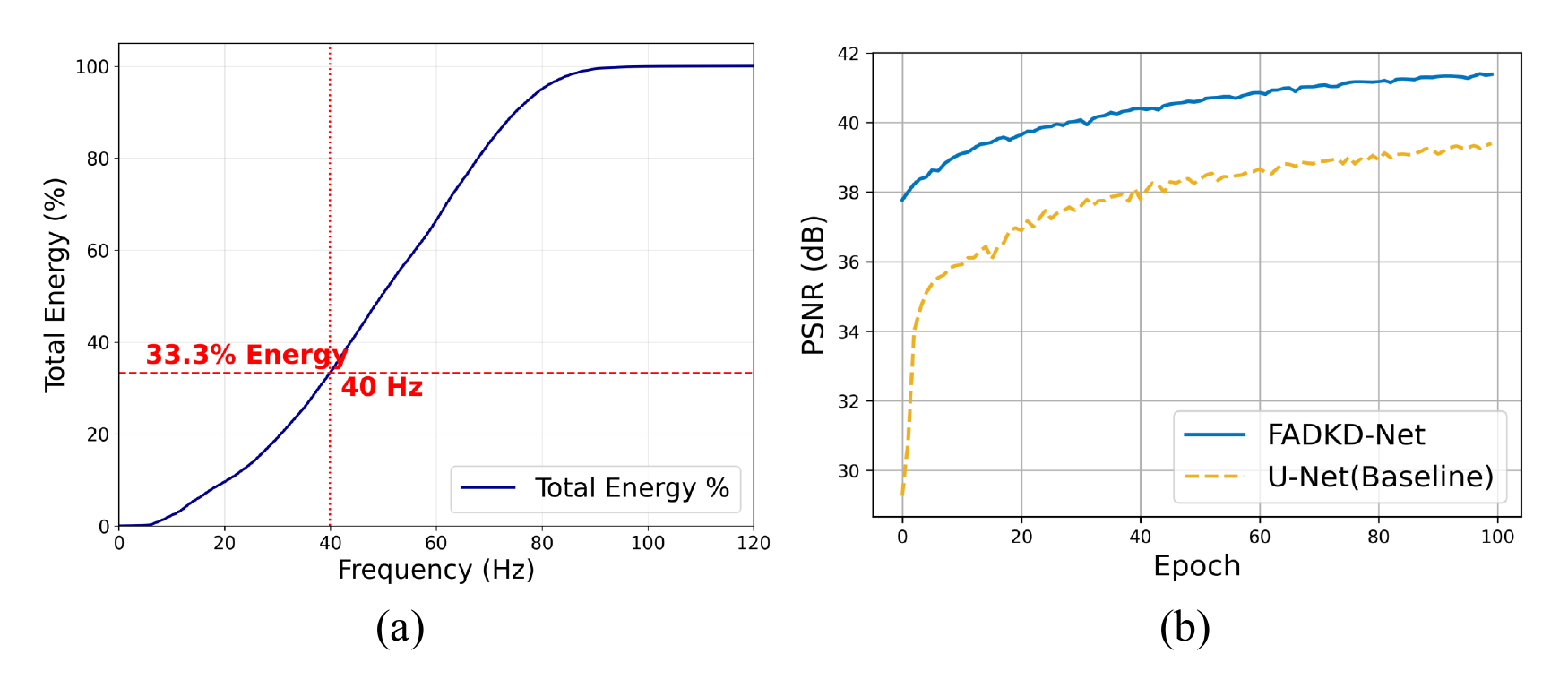}}
	\caption{Frequency-energy spectrum and PSNR evaluation on the Kerry dataset.  (a) The frequency-energy spectrum; (b) PSNR values on the validation sets.}
	\label{fig_kerry_dataset}
\end{figure}

Different from the MAVGL 12 experiment, where the teacher model is trained on synthetic data, the teacher model for the Kerry dataset is trained using another poststack field dataset, namely the New Zealand Parihaka dataset, with a sampling interval of 4 ms. Specifically, a seismic volume with dimensions of $320 \times 416 \times 712$ ($Time \times Crossline \times Inline$) is used for teacher model training, from which 1920 patches of size $64 \times 64 \times 64$ are extracted using a sliding window with a stride of $32 \times 32 \times 32$. A cutoff frequency of 28 Hz is adopted to decouple the seismic features into low- and high-frequency components. Compared with synthetic-to-real knowledge transfer, real-to-real transfer reduces domain discrepancy between the source and target datasets and preserves more authentic geological characteristics. Therefore, the teacher model can provide more reliable prior knowledge for reconstructing the target-domain seismic data.

\begin{figure}[htpb]
	\centering
	{\includegraphics[width=0.99\linewidth]{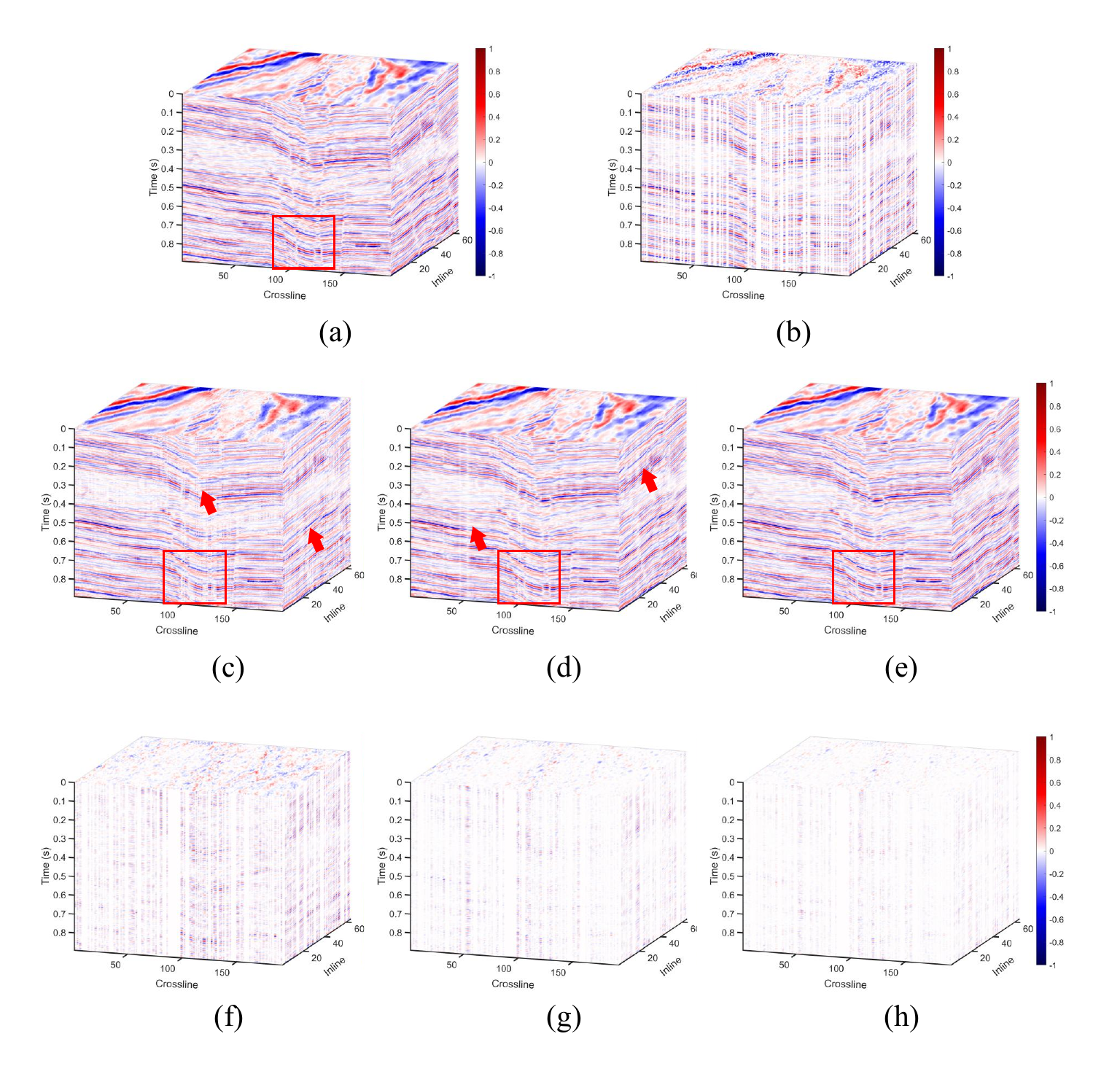}}
	\caption{Interpolation results on the Kerry dataset. (a) Ground truth. (b) Data with 50\% randomly sampled. (c)-(e) Interpolation results corresponding to the DRR, U-Net(Baseline), and FADKD-Net, respectively. (f)–(h) Residual between (a) and (c)-(e), respectively.}
	\label{fig_kerry_results}
\end{figure}

\begin{figure}[htpb]
	\centering
	{\includegraphics[width=0.99\linewidth]{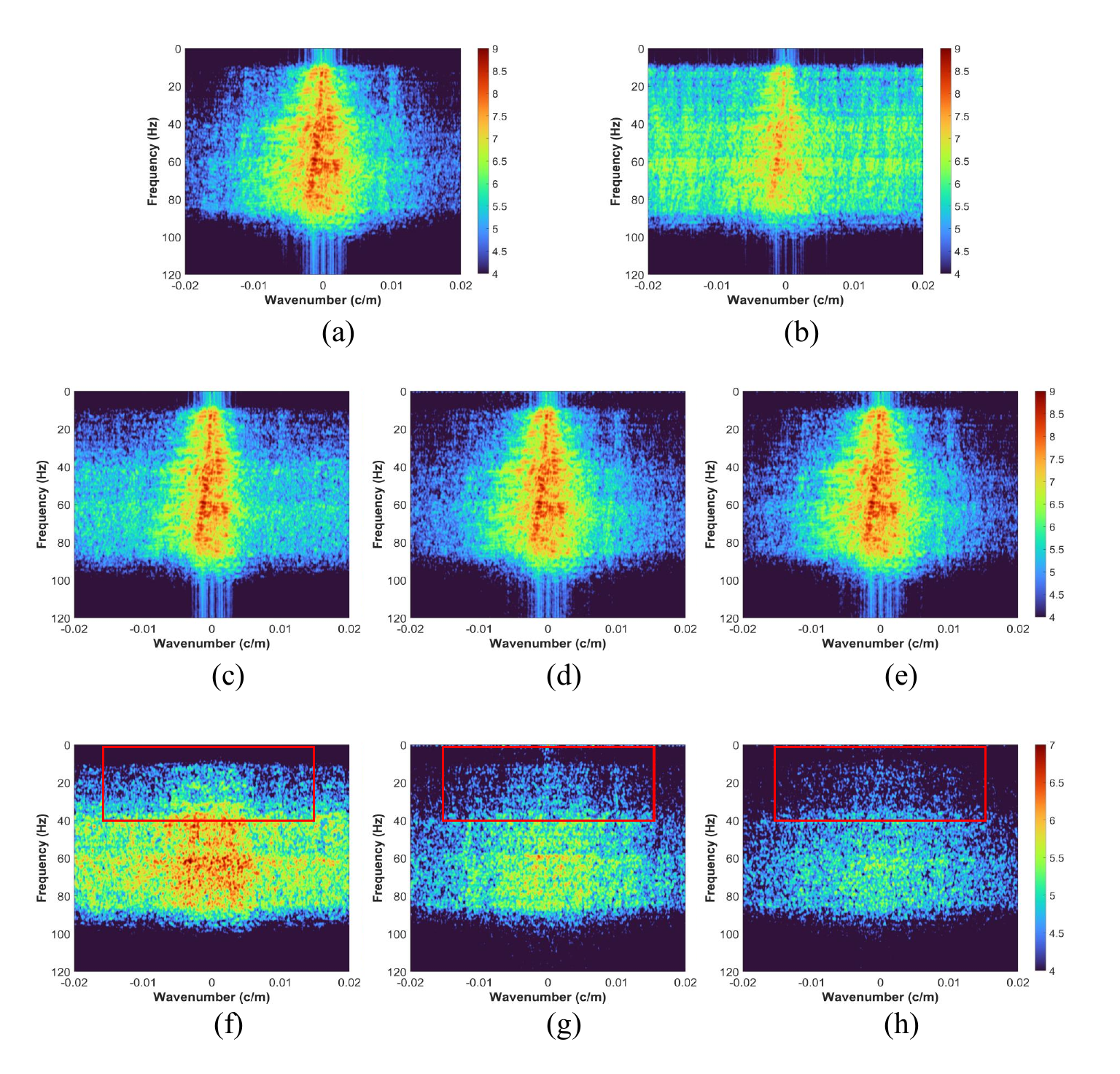}}
	\caption{F-K spectra comparisons of the kerry dataset. (a)-(h) F-K spectra of Figure~\ref{fig_kerry_results}(a)-(h).}
	\label{fig_kerry_fk}
\end{figure}

\begin{figure}[htpb]
	\centering
	{\includegraphics[width=0.99\linewidth]{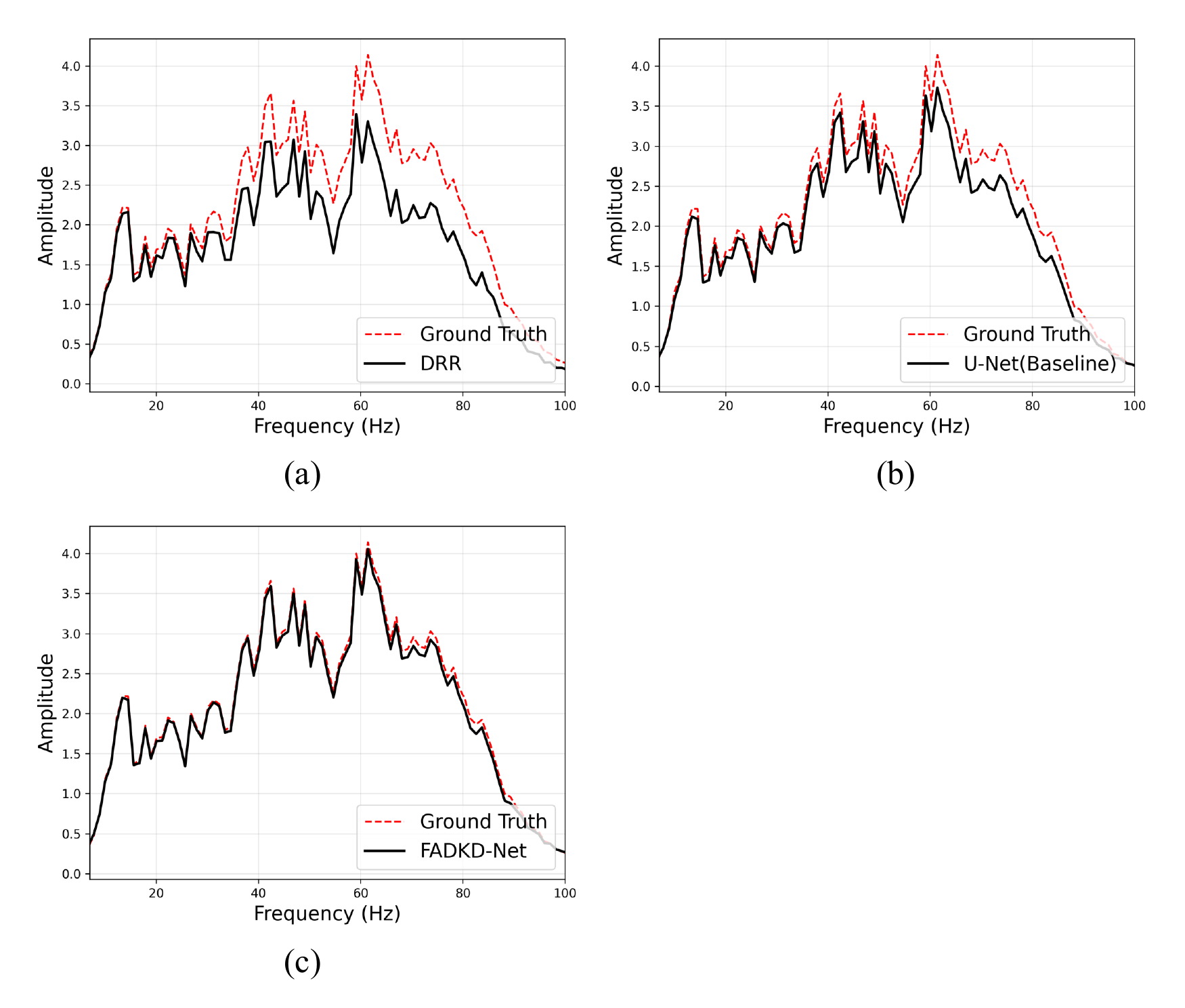}}
	\caption{Spectrum comparisons of the Kerry dataset. (a)-(c) Spectrum results corresponding to the DRR, U-Net(Baseline), and FADKD-Net, respectively.}
	\label{fig_kerry_freq}
\end{figure}

Figure~\ref{fig_kerry_dataset} presents the frequency–energy spectrum of the Kerry dataset and the corresponding PSNR curve on the validation set. According to the energy distribution, the low-frequency cutoff frequency is determined as 40 Hz. Figure~\ref{fig_kerry_results} illustrates the interpolation results. It can be observed that, for this more complex dataset, both DRR (PSNR: 30.89 dB, SSIM: 0.904) and U-Net (Baseline) (PSNR: 36.04 dB, SSIM: 0.972) fail to reconstruct several seismic traces, particularly in the regions indicated by the red arrows. These reconstruction errors are further reflected in the residual maps, where significant signal leakage can be observed. In contrast, FADKD-Net achieves a PSNR of 38.76 dB and an SSIM of 0.985, producing reconstruction results that are closer to the ground truth. This improvement benefits from the incorporation of prior knowledge and the frequency-aware distillation strategy. Specifically, low-frequency knowledge distillation enables FADKD-Net to capture large-scale geological structures and maintain overall seismic event continuity. Meanwhile, high-frequency knowledge transfer improves the recovery of fine-scale variations, resulting in sharper seismic events and more accurate reconstruction of local details, as highlighted in the red box region of Figure~\ref{fig_kerry_results}. 

The F–K spectra shown in Figure~\ref{fig_kerry_fk} further demonstrate the effectiveness of the proposed framework in the frequency domain. Compared with DRR and U-Net (Baseline), the reconstructed spectrum of FADKD-Net exhibits a closer match to the ground truth spectrum, with reduced signal leakage across different frequency regions. The spectrum comparison in Figure~\ref{fig_kerry_freq} provides additional evidence that FADKD-Net achieves more accurate frequency-domain reconstruction over the entire frequency range.

\subsection{Frequency-aware Knowledge Distillation Analysis}
\begin{figure}[htpb]
	\centering
	{\includegraphics[width=0.99\linewidth]{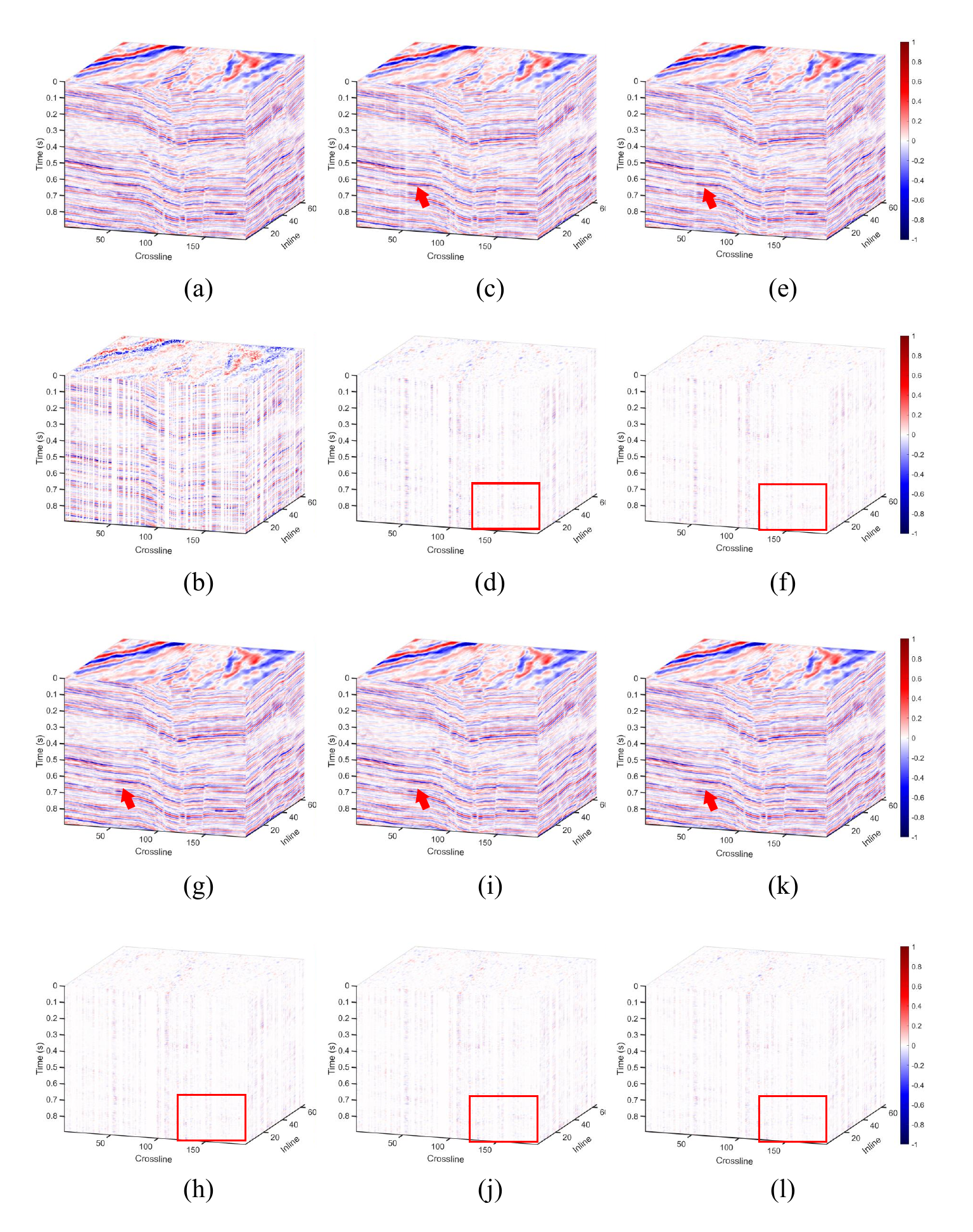}}
	\caption{Interpolation results under different frequency-aware distillation strategies on the Kerry dataset. (a) Ground truth. (b) Data with 50\% randomly sampled. (c), (e), (g), (i), and (k) Interpolation results of Baseline, Baseline-KD, Baseline-KDL, Baseline-KDH, and FADKD-Net, respectively. (d), (f), (h), (j), and (l) Residuals between (a) and the corresponding reconstructions.}
	\label{fig_kerry_abl}
\end{figure}

To validate the contribution of the proposed frequency-aware distillation strategy, ablation experiments are conducted on the Kerry dataset. Specifically, three distillation schemes are compared: U-Net with full-band knowledge distillation (Baseline-KD), U-Net with low-frequency distillation (Baseline-KDL), and U-Net with high-frequency distillation (Baseline-KDH). The corresponding interpolation results are presented in Figure~\ref{fig_kerry_abl}.

\begin{table}[h]
	\centering
	\caption{SSIM and PSNR results for the Kerry dataset with different methods.}
	\label{table_kerry_abl}
	\begin{tabular}{|l|c|c|c|c|c|}
		\hline
		\textbf{Method} & \textbf{U-Net(Baseline)} & \textbf{Baseline-KD} & \textbf{Baseline-KDL} & \textbf{Baseline-KDH} & \textbf{FADKD-Net} \\
		\hline
		PSNR/dB&36.04&37.32&38.03&38.30& \textbf{38.76} \\
		SSIM&0.972&0.979&0.983&0.981& \textbf{0.985} \\
		\hline
	\end{tabular}
\end{table}

(1)	The contribution of full-band distillation. As shown in Figure~\ref{fig_kerry_abl}(c), the Baseline model produces unsatisfactory interpolation results because it is trained only with target source data. Several missing traces remain unrecovered, as indicated by the red arrows. After introducing the teacher model for knowledge distillation, the student model receives additional prior knowledge from the teacher model. Consequently, Baseline-KD achieves improved interpolation results, and some missing traces are successfully reconstructed. However, without frequency-specific guidance, full-band distillation treats all frequency components equally and still struggles to fully recover the missing traces.

(2)	The contribution of low-band distillation. Building upon the full-band distillation, low-frequency distillation is introduced to explicitly guide the learning of large-scale structural information. As shown in Figure~\ref{fig_kerry_abl}(g), incorporating low-frequency guidance enables the model to better preserve the continuity of the seismic events. In particular, the missing traces indicated by the red arrows are reconstructed more accurately compared with Baseline-KD. However, since low-frequency distillation mainly focuses on global structural information, it remains limited in recovering fine-scale details.

(3)	The contribution of high-band distillation. To further enhance the reconstruction of detailed structures, high-frequency distillation is introduced to emphasize fine- scale information. As illustrated in Figure~\ref{fig_kerry_abl}(i), high-frequency guidance significantly improves the recovery of local details, particularly in regions with abrupt variations and complex textures. The missing traces highlighted by the red arrows are reconstructed more completely, with clearer boundaries and sharper structural transitions. Moreover, fewer signal leakages are observed in the red box region of the residual map.

Compared with full-band and low-band distillation, high- frequency distillation provides stronger capability for recovering complex local details. However, high-frequency information alone cannot fully preserve large-scale structural continuity. By combining low- and high-frequency distillation, the model achieves a more balanced reconstruction across different frequency components. The quantitative results in Table~\ref{table_kerry_abl} further support this observation, where the combination of low- and high-frequency distillation achieves the highest PSNR and SSIM values.

\begin{table}[h]
	\centering
	\caption{Training time.}
	\label{table_training_time}
	\begin{tabular}{|l|c|c|c|}
		\hline
		\textbf{Method} & \textbf{U-Net(Baseline)} & \textbf{Teacher model} & \textbf{Student model} \\
		\hline
		MAVGL 12&0.16 h&0.19 h&0.17 h \\
		Kerry&0.29 h&0.47 h&0.43 h \\
		\hline
	\end{tabular}
\end{table}

\begin{table}[h]
	\centering
	\caption{GPU memory occupied by different methods.}
	\label{table_gpu}
	\begin{tabular}{|l|c|c|c|}
		\hline
		\textbf{Method} & \textbf{U-Net(Baseline)} & \textbf{Teacher model} & \textbf{Student model} \\
		\hline
		MAVGL 12&9593 MiB&9593 MiB&16853 MiB \\
		Kerry&9593 MiB&9593 MiB&16853 MiB \\
		\hline
	\end{tabular}
\end{table}

\begin{figure}[htpb]
	\centering
	{\includegraphics[width=0.99\linewidth]{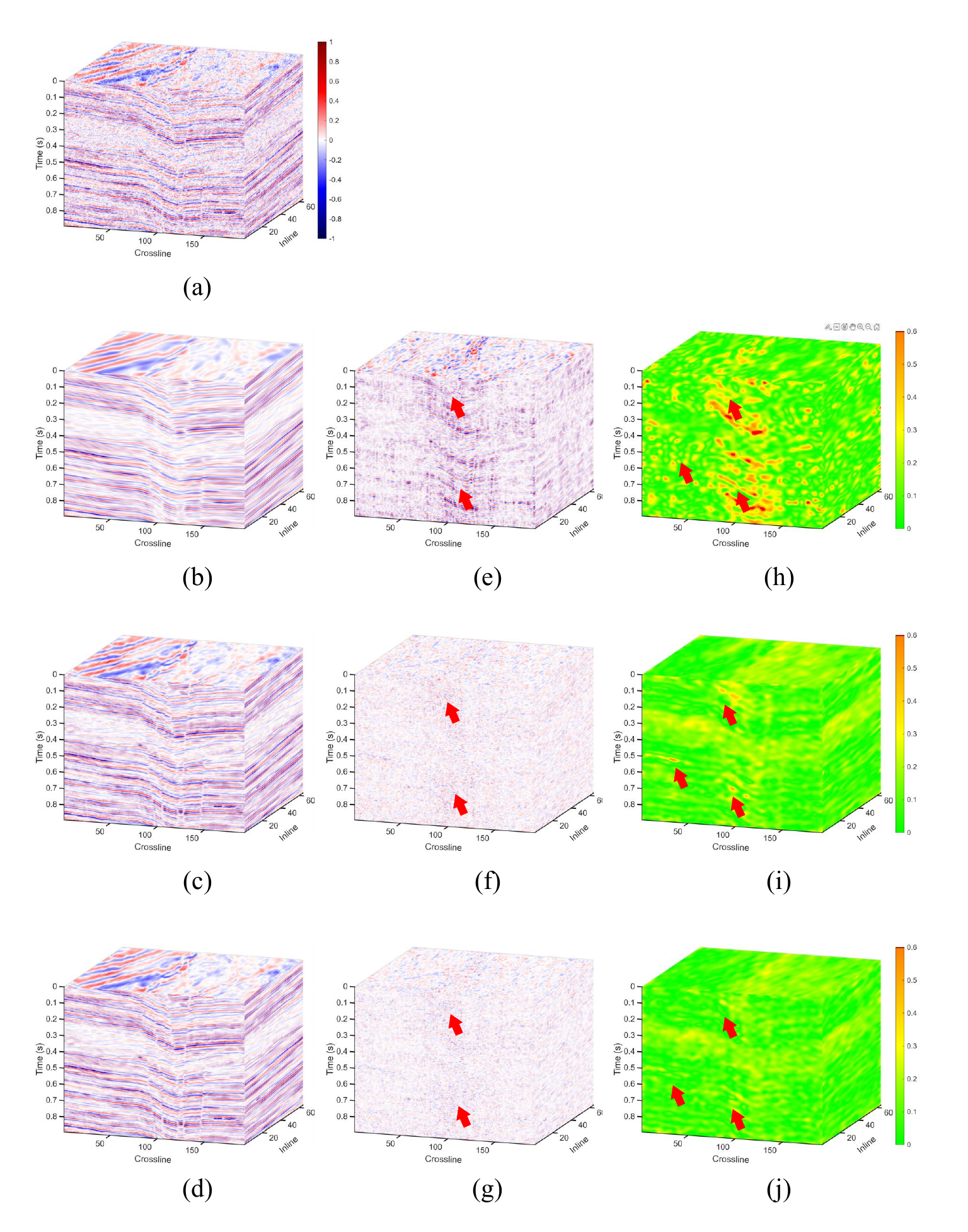}}
	\caption{Denoising results on the Kerry dataset. (a) Noisy data. (b)-(d) Denoising results of DRR, U-Net(Baseline), and FADKD-Net, respectively. (e)-(g) The corresponding residual maps between (b)-(d) and (a). (h)-(j) The corresponding local similarity maps between (b)-(d) and (e)-(g)}
	\label{fig_kerry_dn}
\end{figure}

\subsection{Computational Efficiency Analysis}
To evaluate the computational cost of FADKD-Net, we analyze the training time and GPU memory consumption. All experiments are conducted on the same hardware platform under identical training configurations. The computational comparison is performed between U-Net(Baseline) and the proposed FADKD-Net.

Table~\ref{table_training_time} and Table~\ref{table_gpu} summarize the computational comparison between the two methods. Compared with U-Net, FADKD-Net introduces an additional teacher model and frequency-aware distillation strategy during the training stage, resulting in a moderate increase in computational cost. However, considering the improvements in interpolation performance achieved by the FADKD-Net, the training time and GPU memory consumption remain within an acceptable range.

\section{Discussion}
\subsection{Extension to Denoising Task}
The core idea of FADKD-Net is to learn transferable frequency-dependent representations from high-quality data, which provides potential applicability to other seismic processing tasks. To further investigate this capability, an additional denoising experiment is conducted on the Kerry dataset. Similar to the interpolation task, the denoising task aims to recover reliable seismic representations from degraded observations. Specifically, clean data from the Parihaka dataset are regarded as high-quality observations for constructing the teacher model, while noisy seismic data from the Kerry dataset are used as inputs for training the student model. Since field data inherently contain noise, the loss weights are carefully rebalanced when performing the denoising task \cite[]{ref51}, ${\lambda _1} = 1.5,{\lambda _{low}} = 0.1,{\lambda _{high}} = 0.05$. During training, Gaussian noise is added to the training data to simulate random noise, and the model is trained for 60 epochs. To further evaluate the denoising performance, residual maps and local similarity maps are employed. The local similarity maps are calculated between the denoised results and their corresponding residuals, where values closer to 1 indicate stronger similarity between the recovered signals and residual components, suggesting more severe signal leakage.

The denoising results are shown in Figure~\ref{fig_kerry_dn}. Figure~\ref{fig_kerry_dn}(a) presents the noisy input, from which it can be observed that all methods are capable of suppressing noise to a certain extent. However, as highlighted by the red arrows in the residual maps (second column) and the local similarity maps (third column) of Figure 2, noticeable signal leakage persists in these methods. In contrast, FADKD-Net effectively alleviates this issue, achieving more satisfactory denoising performance while better preserving useful seismic signals. These results further confirm that the proposed framework is not restricted to a specific task but provides a general strategy for seismic data processing.

\section{Conclusion}
We proposed a novel DL framework, named FADKD-Net, for seismic feature learning by integrating dynamic knowledge distillation and frequency-decoupled learning. Specifically, FADKD-Net establishes a teacher–student framework to transfer prior knowledge from high-quality seismic data to degraded observations, while a frequency-aware distillation strategy enables targeted knowledge transfer across low- and high-frequency components. Low-frequency distillation provides structural priors for preserving seismic event continuity, whereas high-frequency distillation promotes fine-scale feature modeling and detail recovery. In addition, cross-domain feature alignment reduces distribution discrepancies across seismic surveys and enhances the transferability of the learned features. Using 3D seismic interpolation as a downstream task, experiments on multiple field seismic datasets demonstrate that FADKD-Net achieves promising performance and robust generalization across different surveys. These results demonstrate the effectiveness of frequency-aware knowledge transfer for seismic feature learning and highlight the potential of FADKD-Net as a general framework for transferable seismic learning. In future work, we plan to apply the FADKD-Net to more seismic applications, such as super-resolution, fault detection, and structural interpretation, to explore its potential as a general framework for transferable seismic learning.

\section{Funding}
This work was supported by the National Natural Science Foundation of China (Grant No. 42574169) and the Deep Earth Probe and Mineral Resources Exploration - National Science and Technology Major Project (Grant No. 2025ZD1007600).

\section{Conflict of Interest}
The authors declare no conflict of interest.

\section{Data and Materials Availability}
Data associated with this research are available and can be obtained by contacting the corresponding author.

% --- Bibliography ---
\bibliographystyle{apalike} % or choose another style like 'apalike', 'ieeetr', etc.
\bibliography{references}

\end{document}